\input harvmac
\input epsf

\newcount\figno
\figno=0
\def\fig#1#2#3{
\par\begingroup\parindent=0pt\leftskip=1cm\rightskip=1cm\parindent=0pt
\baselineskip=12pt
\global\advance\figno by 1
\midinsert
\epsfxsize=#3
\centerline{\epsfbox{#2}}
\vskip 14pt

{\bf Fig. \the\figno:} #1\par
\endinsert\endgroup\par
}
\def\figlabel#1{\xdef#1{\the\figno}}
\def\encadremath#1{\vbox{\hrule\hbox{\vrule\kern8pt\vbox{\kern8pt
\hbox{$\displaystyle #1$}\kern8pt}
\kern8pt\vrule}\hrule}}

\overfullrule=0pt

\noblackbox
\parskip=1.5mm

\def\Title#1#2{\rightline{#1}\ifx\answ\bigans\nopagenumbers\pageno0
\else\pageno1\vskip.5in\fi \centerline{\titlefont #2}\vskip .3in}

\font\caps=cmcsc10

\noblackbox
\parskip=1.5mm



           \def\CO{{\cal O}} 
   
\def\CL{{\cal L}}   
  \def\CD{{\cal D}}


\def\dj{\hbox{d\kern-0.347em \vrule width 0.3em height 1.252ex depth
-1.21ex \kern 0.051em}}

\def\Tr{{\rm Tr\,}}

\def\ket{\rangle}
\def\bra{\langle}

\def\pt{\partial}

\def\Dirac{\,\raise.15ex\hbox{/}\mkern-13.5mu D}
\def\dirac{\,\raise.15ex\hbox{/}\kern-.57em \partial}
\def\aslash{\,\raise.15ex\hbox{/}\mkern-13.5mu A}

\def\shalf{{\ifinner {\textstyle {1 \over 2}}\else {1 \over 2} \fi}} 
\def\sshalf{{\ifinner {\scriptstyle {1 \over 2}}\else {1 \over 2} \fi}} 
\def\sfourth{{\ifinner {\textstyle {1 \over 4}}\else {1 \over 4} \fi}}

\lref\zamo{
A.~B.~Zamolodchikov,
  ``Expectation value of composite field T anti-T in two-dimensional quantum field theory,''
[hep-th/0401146].
}

\lref\zamodos{
F.~A.~Smirnov and A.~B.~Zamolodchikov,
  ``On space of integrable quantum field theories,''
Nucl.\ Phys.\ B {\bf 915}, 363 (2017).
[arXiv:1608.05499 [hep-th]].
}

\lref\cardycorr{
 J.~Cardy,
  ``$T\bar T$ deformation of correlation functions,''
JHEP {\bf 1912}, 160 (2019), [JHEP {\bf 2019}, 160 (2020)].
[arXiv:1907.03394 [hep-th]].

}

\lref\cardy{
J.~L.~Cardy,
  ``Operator Content of Two-Dimensional Conformally Invariant Theories,''
Nucl.\ Phys.\ B {\bf 270}, 186 (1986).
}

\lref\carlip{
 S.~Carlip,
  ``Logarithmic corrections to black hole entropy from the Cardy formula,''
Class.\ Quant.\ Grav.\  {\bf 17}, 4175 (2000).
[gr-qc/0005017].
}

\lref\ridge{
 J.~L.~F.~Barbon and E.~Rabinovici,
  ``Touring the Hagedorn ridge,''
In *Shifman, M. (ed.) et al.: From fields to strings, vol. 3* 1973-2008.
[hep-th/0407236].

J.~L.~F.~Barbon and E.~Rabinovici,
  ``Aspects of Hagedorn holography,''
Les Houches {\bf 87}, 449 (2008).
}

\lref\kut{
 D.~Kutasov and D.~A.~Sahakyan,
  ``Comments on the thermodynamics of little string theory,''
JHEP {\bf 0102}, 021 (2001).
[hep-th/0012258].
}

\lref\us{
 J.~L.~F.~Barbon, C.~A.~Fuertes and E.~Rabinovici,
  ``Deconstructing the little Hagedorn holography,''
JHEP {\bf 0709}, 055 (2007).
[arXiv:0707.1158 [hep-th]].
}

\lref\sheikh{
 F.~Loran, M.~M.~Sheikh-Jabbari and M.~Vincon,
  ``Beyond Logarithmic Corrections to Cardy Formula,''
JHEP {\bf 1101}, 110 (2011).
[arXiv:1010.3561 [hep-th]].
}

\lref\uvc{
S.~Dubovsky, R.~Flauger and V.~Gorbenko,
  ``Solving the Simplest Theory of Quantum Gravity,''
JHEP {\bf 1209}, 133 (2012).
[arXiv:1205.6805 [hep-th]].

 S.~Dubovsky, V.~Gorbenko and M.~Mirbabayi,
  ``Asymptotic fragility, near AdS$_{2}$ holography and $ T\overline{T} $,''
JHEP {\bf 1709}, 136 (2017).
[arXiv:1706.06604 [hep-th]].

L.~McGough, M.~Mezei and H.~Verlinde,
  ``Moving the CFT into the bulk with $ T\overline{T} $,''
JHEP {\bf 1804}, 010 (2018).
[arXiv:1611.03470 [hep-th]].

 A.~Giveon, N.~Itzhaki and D.~Kutasov,
  ``$ T\overline{T} $ and LST,''
JHEP {\bf 1707}, 122 (2017).
[arXiv:1701.05576 [hep-th]].

P.~Kraus, J.~Liu and D.~Marolf,
  ``Cutoff AdS$_{3}$ versus the $ T\overline{T} $ deformation,''
JHEP {\bf 1807}, 027 (2018).
[arXiv:1801.02714 [hep-th]].

J.~Cardy,
  ``The $ T\overline{T} $ deformation of quantum field theory as random geometry,''
JHEP {\bf 1810}, 186 (2018).
[arXiv:1801.06895 [hep-th]].

 O.~Aharony, S.~Datta, A.~Giveon, Y.~Jiang and D.~Kutasov,
  ``Modular invariance and uniqueness of $T\bar{T}$ deformed CFT,''
JHEP {\bf 1901}, 086 (2019).
[arXiv:1808.02492 [hep-th]].

}

\lref\difran{
P.~Di Francesco, P.~Mathieu and D.~Senechal,
``Conformal Field Theory,''
doi:10.1007/978-1-4612-2256-9. }

\lref\negro{
A.~Cavaglià, S.~Negro, I.~M.~Szécsényi and R.~Tateo,
  ``$T \bar{T}$-deformed 2D Quantum Field Theories,''
JHEP {\bf 1610}, 112 (2016).
[arXiv:1608.05534 [hep-th]].

}

\lref\banksofer{
 O.~Aharony and T.~Banks,
  ``Note on the quantum mechanics of M theory,''
JHEP {\bf 9903}, 016 (1999).
[hep-th/9812237].
}

\lref\gross{
 D.~J.~Gross, J.~Kruthoff, A.~Rolph and E.~Shaghoulian,
  ``$T\overline{T}$ in AdS$_2$ and Quantum Mechanics,''
Phys.\ Rev.\ D {\bf 101}, no. 2, 026011 (2020).
[arXiv:1907.04873 [hep-th]].

 D.~J.~Gross, J.~Kruthoff, A.~Rolph and E.~Shaghoulian,
  ``Hamiltonian deformations in quantum mechanics, $T\bar T$, and SYK,''
[arXiv:1912.06132 [hep-th]].

}

\lref\fubini{
V.~de Alfaro, S.~Fubini and G.~Furlan,
  ``Conformal Invariance in Quantum Mechanics,''
  Nuovo Cim.\ A {\bf 34}, 569 (1976).
  }
  
  \lref\fubinir{
  S.~Fubini and E.~Rabinovici,
  ``Superconformal Quantum Mechanics,''
Nucl.\ Phys.\ B {\bf 245}, 17 (1984).
}

\lref\ivanov{
V.~Akulov and A.~Pashnev,
``Quantum Superconformal Model In  (1,2) Space,''
Theor.\ Math.\ Phys.\  {\bf 56}, 862-866 (1983).
}
  
\lref\thirring{
W. Thirring, ``Quantum Mathematical Physics: Atoms, Molecules and Large Systems", Springer (2002).}




\baselineskip=15pt

\line{\hfill IFT-UAM/CSIC-20-32}

\vskip 0.5cm

\Title{\vbox{\baselineskip 12pt\hbox{}
 }}
{\vbox {\centerline{ Remarks On The Thermodynamic Stability
 }
\vskip10pt
\centerline{Of $T{\bar T}$ Deformations}
}}
\vskip 0.5cm

\centerline{$\quad$ {\caps
Jos\'e L.F. Barb\'on$^\dagger$
 and
Eliezer Rabinovici$^{\star}$
}}
\vskip0.5cm

\centerline{{\sl  $^\dagger$ Instituto de F\'{\i}sica Te\'orica IFT UAM/CSIC }}
\centerline{{\sl  C/ Nicol\'as Cabrera 13,
 Campus Universidad Aut\'onoma de Madrid}}
\centerline{{\sl  Madrid 28049, Spain }}
\centerline{{\tt jose.barbon@csic.es}}

\vskip0.1cm

\centerline{{\sl $^\star$
Racah Institute of Physics, The Hebrew University }}
\centerline{{\sl Jerusalem 91904, Israel}}
\centerline{{\sl and}}
\centerline{{\sl Institut des Hautes Etudes Scientifiques}}
\centerline{{\sl  91440 Bures-sur-Yvette, France}} 
\centerline{{\tt eliezer@vms.huji.ac.il}}

\vskip1cm

\centerline{\bf ABSTRACT}

 \vskip 0.3cm
We point out that negative specific heat at high energies is a characteristic feature of  many $T{\bar T}$ deformations, both in the original $d=2$ case and in $d=1$ quantum mechanical  cousins. This note is a contribution to the memorial volume in honor of P. G. O. Freund.

 \noindent

\vskip 1.5cm

\Date{April 2020}

\vfill

\vskip 0.1cm




\baselineskip=15pt

\newsec{Introduction}

\noindent

Peter Freund's research reflects a lively fascination with boundaries.  
From time to time physicists are faced with boundaries, be they hard such as the
speed of light in empty space,  a rigorous proof that internal and
space-time symmetries cannot be mixed in a non-trivial manner, or be they warning ``border ahead" type posts such as $\hbar$ and the Planck mass. 
When facing bounds physicists try to fully understand their implications but also search for all possible ways to undermine them. They try to find ways around them, they gauge anomalous symmetries, they would do anything to just be able to take a forbidden glance of what
may lie behind the boundaries.  These excursions are many times doomed and heavily punished, they are however 
 handsomely rewarded in some rare cases. Be that as it may, the temptations are out there and must be faced, not resisted. 
 
In this note we discuss circumstances where two such bounds play a role. The main one being the Wilsonian type of bound on the allowed classes of  deformations one may  add
to a well defined theory and remain within the realm of well defined theories. The  other bound is a  potentially maximally allowed temperature --the Hagedorn temperature.
Let us start by considering the issue which technically goes under the warning ``do not add
irrelevant operators".  
Given a well defined field theory one is allowed to add to it truly marginal as well as relevant operators, if they exist in the theory. Adding  irrelevant
operators seems to endanger the proper Ultra Violet (UV) properties of the system. This needs qualifications. 
Given a well defined theory in the UV, it could be asymptotically free {\it ab initio} or 
we may explicitly add to it a relevant operator. Then follow the enforced  trajectory of the system all the way
to the Infrared (IR), where the theory is actually well described by a conformal  system (which could be
trivial) plus  irrelevant operators. That theory has a UV completion: it is the starting point of the flow. 
However there is an important price to pay for this UV completion, as one returns to the UV more and more degrees of freedom are generically added
the system. The IR theory plus an irrelevant operator  does not generically make sense in the UV on its own.

An strategy to overcome the challenge issued by Wilson's dogma is to add to a well defined
theory a very special operator, one that would ensure that the total number of degrees of freedom  in its UV completion is identical to the number of the initial degrees of freedom. For example,  in two dimensions,  
deforming a Conformal Field Theory (CFT)  by a truly marginal operator ensures the conservation of the number of degrees of freedom, as measured by the central charge. The crucial question is whether a special  {\it irrelevant} operator can achieve the same feat.  A proposal along these lines is the so-called  $T{\bar T}$ deformation introduced by A.B. Zamolodchikov \refs{\zamo,\zamodos,\negro}. In this construction, a fine-tuned irrelevant operator is added to a CFT in such a way that the theory remains integrable, and the spectrum can be exactly solved as a function of the deformation parameter. In other words, no extra UV degrees of freedom or dynamics are apparently needed, beyond the information already present in the low energy theory.  The operator that performs this feat is roughly a  `quasiclassical' operator product, that is to say, it is a composite operator appearing in a regular OPE contribution 
\eqn\ttbar{
  \sum_{\alpha\beta} C_{\alpha\beta} {\cal T}_\alpha (y) {\cal T}_\beta (x)  \sim T{\bar T}(x)  + {\rm derivatives}  \;.
}
In this expression, the $C_{\alpha\beta}$ coefficients are carefully chosen to cancel the  short-distance singularities of the ${\cal T} \,{\cal T}$ products  in the OPE limit $y\rightarrow x$. Let us further suppose that  the expectation value of the right hand side in energy eigenstates  has no dependence on $(x-y)$ (up to possible contact terms),  so that they can be evaluated using clustering, as quadratic functions of expectation values of the ${\cal T}_\alpha$  operators.  In $d=2$, one can find operators exhibiting these properties among the components of the energy-momentum tensor, which ultimately leads to expectation values in energy eigenstates of the form 
\eqn\ex{
\left\langle E_n |\,T{\bar T} \,|E_n  \right\rangle \sim \pt_\lambda E_n \big |_{\lambda =0} \sim O(E_n^2 )\;,
\;}
upon adding  the dimension-four operator $T{\bar T} $ to the action, with coupling $\lambda$.

The spirit of the $T{\bar T}$ deformation is to go beyond leading order and use the differential equation \ex\ to define a spectral flow at finite values of the deformation parameter $\lambda$. Focusing, for simplicity, on the zero-momentum sector of a theory defined on a spatial circle of radius $R$, one finds an explicit spectral flow given by 
\eqn\ttbarzm{
 F_n =  {1\over 4\pi R \lambda} \left(\sqrt{1+8\pi \lambda R E_n } -1\right)
\;,
}
where we use the notation $F_n = E_n (\lambda)$ for the deformed spectrum and $E_n = E_n (0)$  for the undeformed one. The dimensionless deformation parameter is    $\lambda = (\ell /2\pi R)^2$, in terms of the characteristic length scale $\ell$,  introduced by the perturbing operator.

The natural question is whether we actually succeeded in the Wilsonian challenge. Is  the deformed theory  well defined  after all?  Just from the properties of the spectrum, we notice some general constraints. If $\lambda <0$  the spectrum becomes complex above some high-energy threshold, indicating some sort of short-distance instability and a  failure of the Wilsonian challenge. For positive values of $\lambda$ and a negative vacuum energy in the undeformed theory, such as one provided by a bosonic Casimir energy, the low-energy spectrum becomes unstable for $\lambda > 1/8\pi R |E_0| $, but no blatant pathologies occur in the $0<\lambda < 1/8\pi R |E_0| $  window. Considering for instance  a CFT with central charge $c$  on the circle, its  ground state energy is $E_0 = -c/12 R$,  and we learn that the deformation parameter must lie in the interval
\eqn\intla{
0 < \lambda < {3 \over 2\pi c}\;,
}
which is quite small for large values of the central charge. This bound translates into a minimal hierarchy between the Wilsonian length scale $\ell$ and the size of the circle:
\eqn\hier{
{\ell 
\over 2\pi R} < \sqrt{3 \over 2\pi c}\;.
}
For systems with non-negative vacuum energy, such as supersymmetric ones, there is no immediate constraint on the size of  $\lambda$, provided it is positive.

In general if the original theory has some symmetry which manifests itself in the degeneracy of the energy levels this degeneracy is maintained by
the spectral flow. This ensures that one can redefine such a symmetry also in the deformed system.
We point out two consequences of this observation.
First, if  one starts off with a theory for which the Hamiltonian must vanish when acting on all states, the deformed Hamiltonian will have the same property with
the same wave functions as the original theory. Thus such deformation of a topological theory retains its topological features as would those of a system
obeying the Wheeler-De Witt equation.
Second,  if one starts off with a supersymmetric theory the deformed theory will conserve the supersymmetric spectrum and moreover if the supersymmetry was not
spontaneously broken initially, there will be no spontaneous breaking of supersymmetry also in the deformed symmetry. This can be seen in two ways.
First the value zero for the energy is a fixed point of the flow, and secondly the wave functions are preserved and thus also their normalizability properties.
A zero energy, normalizable wave function, will remain a normalizable zero energy state of the deformed theory. This works also in the opposite direction. 
Supersymmetry cannot  be generated and a spontaneous breaking cannot  be annulled by the deformation.

 A notable feature of \ttbarzm\ is the high-energy form of the flow, which gives $R F_n \approx \sqrt{RE_n /2\pi \lambda}$. This means that the standard Wilsonian density of states of a two-dimensional theory  is turned into a `stringy' density of states. To see this, consider the function ${\cal N}_0 (E)$, which counts the number of energy eigenstates below $E$. If the  undeformed theory is defined by a UV fixed point of central charge $c$, Cardy's formula \refs\cardy\ gives the leading high-energy asymptotic form ${\cal N}_0 (E) \approx \exp(S_C (E))$, where 
 \eqn\card{
 S_C (E) = 2\pi \sqrt{ {c\over 3} E\,R}\;
 }
 is the so-called Cardy entropy. 
 
  For any flow $E\rightarrow E(\lambda) = F(E)$ without level crossings, such as \ex,   the level-counting function after deformation ${\cal N}(F)$ satisfies ${\cal N}(F(E)) = {\cal N}_0 (E)$. Therefore, expressing \card\ in terms of $F$ we find  an asymptotic high-energy spectrum of Hagedorn type 
\eqn\twodim{
{\cal N} (F)  \sim e^{F/T_s}\;,
}
with  the effective Hagedorn temperature  
\eqn\hag{
T_s = \Lambda_\ell\, \sqrt{ 3\over 2\pi c}
\;, }
where we have defined the energy scale $\Lambda_\ell = 1/\ell$ representing the naive Wilsonian cutoff of the undeformed theory. In deriving \hag\ we have used  the  previous definition $\lambda = (\ell / 2\pi R)^2$ of UV length scale $\ell$.
It is interesting to notice that, for large values of the central charge, the Hagedorn temperature is much smaller than the naive Wilsonian cutoff of the theory $\Lambda_\ell$. On the other hand, the 
 squeezing  of the spectrum only turns on at $F$-energies larger than 
 \eqn\squez{
 F_s = \Lambda_\ell \, { 2\pi R \over \ell} \gg \Lambda_\ell\;, 
 }
  which becomes the relevant energy threshold in the deformed theory.  Then, we have a bound on the ratio $F_s /T_s$ coming from the ground state stability condition \intla: 
 \eqn\boundd{
 {F_s \over T_s} = \sqrt{2\pi c \over 3\lambda} > {2\pi c \over 3} \;.
 }
 In this respect, for the model to resemble a critical string theory we need to push the deformation parameter towards its upper bound and keep the central charge small. Otherwise $T_s \ll F_s$, i.e.  the Hagedorn temperature is hierarchically smaller than the squeezing   scale. 
 
The emergence of a Hagedorn spectrum ultimately puts into question the physical status of the deformed theory as a standard field theory. Perhaps it is better interpreted as some kind of gravitational theory or effective string theory \refs\uvc.  Certainly, the occurrence  of a Hagedorn density of states questions the existence of standard notions of  local observables \refs\banksofer\ (see however \refs\cardycorr). 

In this note we make two observations. The first is that the Hagedorn phase of the $d=2$ $T{\bar T}$ deformation is ultimately thermodynamically unstable when the seed theory is conformal. This means that a UV completion into a thermodynamically stable phase will tend to hide the Hagedorn phase behind the latent heat of a first-order phase transition.  Our second observation is that quantum mechanical  (QM) versions of the $T{\bar T}$ story (formally the $d=1$ case) also have a tendency to show the same pattern: if the deformation is defined at all at high energies, it tends to produce a high-energy accumulation of levels, resulting in  a negative specific heat. This suggests that the deformed QM  theory is non-local in a strong sense.

\newsec{Thermodynamic Instability Of $T{\bar T}$-deformed CFTs}

\noindent

When confronted with a Hagedorn spectrum at high energies, a natural question is whether the Hagedorn temperature is actually `limiting' in the thermodynamical sense, or rather suggests the existence of a first-order phase transition. To settle this question one examines the microcanonical entropy function, defined by
\eqn\micen{
S(E) = \log\,{\cal N}(E)\;,
}
with ${\cal N}(E)$ as the level-counting function which counts the number energy eigenstates $E_\alpha$ below a given energy $E$, 
\eqn\levelc{
{\cal N}(E) = \Tr \,\theta (E-H) = \sum_\alpha \theta(E-E_\alpha) \;,
}
where $\theta$ stands for the Heaviside step function and we assume discreteness of the energy spectrum. The level counting function is related to the density of states $\Omega(E)$ by the formula
\eqn\densi{
\Omega(E) = {d {\cal N} \over dE} = \Tr \,\delta(E-H) = \sum_\alpha \delta (E-E_\alpha) \;.}
The density of states can be computed as an inverse Laplace transform of the canonical partition function: 
\eqn\can{
Z(\beta) = \Tr \,e^{-\beta H} = \int dE \,\Omega(E) \,e^{-\beta E} \;, \qquad \Omega (E) = \int_\Gamma {d\beta \over 2\pi i} e^{\beta E} \,Z(\beta)\;,
}
where the contour $\Gamma$ runs parallel to the imaginary axis, to the right of all singularities of $Z(\beta)$. Approximate evaluations of  $\Omega(E)$ through \can\ often give smooth approximations of $\Omega(E)$ and thus $S(E)$. 
From these smooth thermodynamical functions we may calculate the microcanonical temperature \foot{We chose to define the entropy \micen\  with maximal coarse-graining in energy (cf. for instance  \refs\thirring), in order to have the simplest mathematical relation to the density of states. Other definitions are possible,  where one replaces $\theta(E-H)$ with a  window function  supported on a narrow energy interval around $E$, with width $\Delta_E$. For a sufficiently small $\Delta_E$, we can then approximate the band dimension as ${\cal N}_\Delta \approx \Delta_E \,\Omega (E)$ and define a band-microcanonical entropy $S_\Delta (E) = \log {\cal N}_\Delta$.  }
\eqn\mictem{
T(E) = \left({\pt S \over \pt E}\right)^{-1}\;.
}
The monotonicity of $T(E)$ determines the thermodynamic stability.  Namely the sign of the specific heat coincides with the sign of $dT/dE$. 
A monotonically increasing $T(E)$ gives positive specific heat and a thermodynamical equivalence between canonical and microcanonical ensembles. On the other hand, if a system with negative specific heat is  embedded in a UV completion with positive specific heat, one expects a first-order phase transition as a function of the temperature, with a critical temperature somewhat below the maximum of the $T(E)$ function. In this situation one has a sort of `Maxwell construction' whereby the system jumps between the two thermodynamically stable bands with a finite latent heat (cf. Figure 1).

  \bigskip
\centerline{\epsfxsize=0.7\hsize\epsfbox{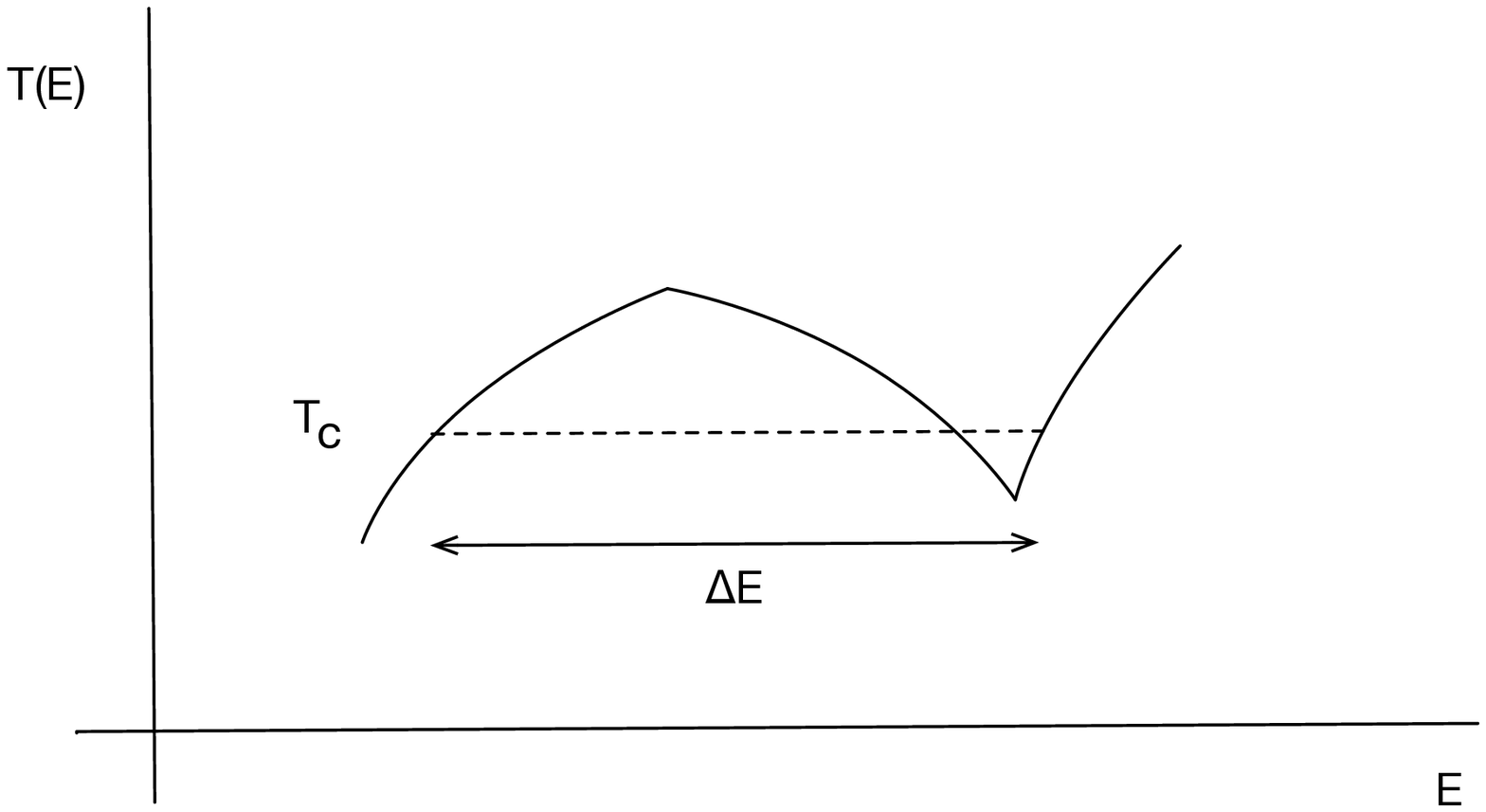}}
\noindent{\ninepoint\sl \baselineskip=2pt {\bf Figure 1:} {\ninerm
Schematic representation of a system with a band of negative specific heat (decreasing microcanonical temperature $T(E)$) bounded by standard bands of positive specific heat. If one uses the temperature as control parameter, the system minimizes free energy by jumping to the high-energy phase at a critical temperature $T_c$. The energy jump $\Delta E$ defines the latent heat of the first-order phase transition.
}}
\bigskip

A large class of systems have  high-energy densities of states of the form
\eqn\genp{
\Omega(E) = {A \over \mu}  \,{e^{S_\mu (E)} \over S_\mu (E)^\delta} \left(1 + O(1/S_\mu)\right)
\;,
}
where $A$ and $\delta$ are positive constants and  $S_\mu (E) = (E/\mu)^\alpha$ with $\mu$ some pivot energy scale and $\alpha >0$. For any such system, one finds 
\eqn\genn{
{\cal N}(E) = {A \over \alpha}  \,{e^{S_\mu (E)} \over S_\mu (E)^\gamma} \left(1+ O(1/S_\mu)\right)\;,\qquad \gamma = \delta + 1-{1\over \alpha}\;,
}
and the microcanonical temperature admits the high energy expansion \foot{The distinction between the exponents $\gamma$ and $\delta$ is largely based on our definition of entropy in \micen, which we chose to be independent of any arbitrary band-binning $\Delta_E$. Should we use a narrow-window definition, with $S_\Delta (E) \approx \log(\Delta_E \Omega(E))$,  the role of $\gamma$ below would be played by the parameter $\delta$. This freedom of definition will have no effect on our qualitative conclusions, which shall depend only on the positive sign of either $
\gamma$ or $\delta$.}
\eqn\microt{
T(E) = {\mu \over \alpha} \left({E \over \mu}\right)^{1-\alpha} \left( 1 + \gamma \left({\mu \over E} \right)^{\alpha} + \dots \right)\;.
}
The specific heat is proportional to $1-\alpha$, so that any system with $\alpha <1$ is thermodynamically stable at high energies, a characteristic example being relativistic QFT in $d$ spacetime dimensions, which has $\alpha = (d-1)/d$. In particular $\alpha =1/2$ for $d=2$, which implies $\gamma = \delta -1$ in this case.\foot{The case of quantum mechanics, formally corresponding to $d=1$, has $T(E) \propto E$ at high energies (cf. Appendix C).} Conversely, Schwarzschild black holes in $d$ spacetime dimensions have $\alpha = (d-2)/(d-3) >1$, a characteristic example of a thermodynamically unstable system. 

The marginal case, $\alpha=1$, is  the Hagedorn spectrum with Hagedorn temperature $T_s = \mu$.   Its thermodynamic stability  depends on the sign of the subleading parameter $\gamma$, which in this case (and only in this case) coincides with $\delta$. Namely a positive (negative) value of $\gamma$ corresponds to a negative (positive) specific heat in the Hagedorn band. 
In a system possessing a Hagedorn spectrum with positive specific heat, the Hagedorn temperature is physically a maximal temperature and the given description is in principle self-consistent. On the other hand, with $\gamma <0$ the Hagedorn temperature is slightly surpassed and approached from above at high energies. As indicated above, the resulting physical picture depends very sensitively on whether there exist UV completions with positive specific heat (cf. \refs{\ridge, \us} for a review adapted to the context of Hagedorn phases).

From the definition \micen\  and the condition of regular spectral flow $ {\cal N}(F(E)) = {\cal N}_0 (E)$, we deduce
the functional relation $
 S(F(E)) = S_0 (E) $, which implies
\eqn\mict{
T(F) = F' (E) \,T_0 (E)
}
for the microcanonical temperatures (here the prime denotes derivative with respect to $E$). 

In computing $T(F)$ from \mict, we focus on a unitary CFT as the undeformed system. It turns out that, in estimating the high energy behavior of $T(F)$, it is essential to keep track of the logarithmic corrections to Cardy's formula \card. Following   \refs{\carlip, \sheikh}, we have 
\eqn\cardy{
S_0 (E) = S_C (E) - \gamma \,\log S_C (E) + O(E^0) 
\;,
}
as $E\rightarrow \infty$, with $S_C (E) = 2\pi \sqrt{cE/3} $ and $\gamma$  a positive constant. Here and in what follows we choose units so that $R=1$.  

For simplicity, we begin by focusing on the zero-momentum sector, for which the deformation map \ttbarzm\  is the simplest. When the undeformed system is a compact, unitary CFT with discrete spectrum of conformal dimensions, this  corresponds to  the choice 
\eqn\gammados{
\gamma \big |_{P=0} =2\;,
} 
 a result which we review in the Appendix 1, where we also explain how $\gamma$ gets modified in some examples of non-compact CFTs. At any rate, we keep an indefinite value of the coefficient $\gamma$ in what follows.  The $P=0$  microcanonical temperature of the undeformed theory takes the form 
\eqn\cardlog{
T_0 (E) \big |_{P=0}  = T_C (E) \left(1+ {\gamma \over S_C (E)} + O(1/S_C^2) \right) \;,
}
at energies $E\gg 1$, where 
\eqn\mictcero{
T_C (E) = {3 \over 2\pi^2 c} \,S_C (E)\;.
}

From the spectral flow formula  \ttbarzm\ we obtain  
\eqn\efeprima{
F'(E) = {1 \over \sqrt{1+ 8\pi \lambda  E}} = {1\over 1+ 4\pi \lambda F}  = {F_s \over 2F} \left(1+ {F_s \over 2F}\right)^{-1}\;. 
}
Finally, inverting the spectral flow to write all expressions in terms of $F$, 
\eqn\invert{
E = F + 2\pi \lambda\,F^2 = {F^2 \over F_s} \left(1 + {F_s \over F} \right) \;, \qquad S_C (E) = {F \over T_s} \sqrt{1 + {F_s \over F}}\;, 
}
we are led   to the final result   
\eqn\finale{
T(F) \big |_{P=0}  = T_s {\sqrt{1+ {F_s \over F}} \over 1+ {F_s\over 2F}} \,\left(1+ {\gamma T_s \over F} {1\over \sqrt{1+{F_s \over F}}} + O(T_s^2 /F^2) \right)\;,
}
which can be expanded   for $F\gg F_s$ to obtain 
\eqn\finolo{
T(F) \big |_{P=0} = T_s \left( 1 + \gamma{ T_s \over F} - {F_s^2 \over 8F^2} - {\gamma \over 2} {T_s F_s \over F^2} + O(T_s^2/F^2) \right)\;.
}
In this expression, we have written the second-order corrections hierarchically, taking into account that $T_s < F_s$ and the fact that we may even have $T_s \ll F_s$ for large values of the central charge (cf. Eq. \boundd). In any case, we stress that \finolo\ applies for $F\gg F_s$, which is the region where the spectral squeezing takes place. For $F<F_s$ the microcanonical temperature is well-approximated by the CFT one \cardlog. 

 The dominant correction to the constant Hagedorn temperature is linear in $1/F$ and induces a decreasing microcanonical temperature, i.e. a negative heat capacity. We notice that this term comes from  the logarithmic correction to the Cardy formula. Had we ignored this term, the corrections of order $F_s^2 /F^2$, coming entirely from the spectral deformation \ttbarzm, would have produced a positive specific heat. In fact, for the case of $\gamma = O(1)$ and large central charge,  a transient opens up  in the interval $F_s \ll F \ll O(F_s^2 /T_s) \sim O(c \,F_s)$, where the specific heat is indeed positive. 
  As it usually happens with thermodynamically unstable Hagedorn spectra, the concrete   monotonicity of $T(F)$ is very mild at high energies, so that it could be affected by  slight modifications of the assumptions about the high-energy behavior, such as the location and size of  threshold corrections from a concrete proposal of UV completion (cf. \refs{\kut, \us} for considerations along these lines in the context of `Little String Theories'). 

It would be interesting to extend this analysis to the full spectrum, rather than the $P=0$ sector alone. The full spectral flow at arbitrary momentum is given by (again in $R=1$ units) 
\eqn\fulls{
F = {1\over 4\pi \lambda }  \left[\sqrt{1+ 8\pi \lambda E + 16\pi^2  \lambda^2 P^2} -1 \right] \;.
} 
For large momentum, of order $P \sim E$, we have $F\sim E$ at large energies. Therefore,  the deformation of the spectrum is less efficient for $P\neq 0$ and we expect the Hagedorn behavior to be dominated by the $P\sim 0$ sector. An explicit computation which upholds this  conclusion is explained in Appendix 2. In particular, it is found that the microcanonical temperature function has the form 
\eqn\newt{
T(F) = T_s \left( 1 + \gamma'  {T_s \over F} + \dots \right) \;, 
}
where $\gamma' $ is a positive constant which equals $3/2$  when the undeformed CFT is `compact', in the sense of having a discrete spectrum of conformal dimensions,  and equals $(3+N)/2$ when the CFT contains $N$ additional  non-compact free bosons.  Therefore, summing up all the momentum sectors amounts to a rescaling of the $P=0$ heat capacity which does not affect its sign.

\newsec{Thermodynamical Instability Of Deformed QM}

\noindent

Quantum Mechanics (QM), formally  a field theory in $d=1$, may  not be such a good toy model for the Wilsonian challenge, but  it offers the simplest arena to study the detailed phenomenology of spectral flows \refs\gross. In particular, any power of the Hamiltonian serves as a $d=1$ analog of the $T{\bar T}$ operator, since $H$ is an effectively `classical' operator when evaluated in states of definite energy. 
We can offer a heuristic derivation of this fact by considering  a quantum mechanical model with action
\eqn\ac{
S_\lambda = \int dt \,\CL_\lambda (\phi, {\partial_t \phi})\;.
}
Let us suppose that near $\lambda =0$ we have 
$$
S_\lambda \approx S_0 + \lambda \int dt \,\CO_0(t)\;,
$$
with $\CO_0$ an  operator local in time.  Consider now the following formal path-integral expression for the energy eigenvalues as a function
of $\lambda$:
$$
E_n (\lambda) = \bra n | H_\lambda | n \ket = -{1\over {\rm Vol} ({\bf R})}  \log\;\int [\CD \phi]_n e^{-S_\lambda} 
\;,$$
where $H_\lambda$ is the exact $\lambda$-dependent Hamiltonian operator, ${\rm Vol}({\bf R})$ stands for the formal volume of the Euclidean time line and the path integral measure is defined with appropriate boundary conditions to capture the $n$-th energy eigenstate. These eigenstates have an implicit dependence on $\lambda$. However,  the Feynman--Hellmann theorem implies 
$$
\pt_\lambda E_n( \lambda) = \bra n | \partial_\lambda H_\lambda | n \ket
$$
because, for normalized states $\bra \psi_\lambda | \psi_\lambda \ket =1$ we have $\bra \pt_\lambda \psi_\lambda | \psi_\lambda \ket + \bra \psi_\lambda | \pt_\lambda \psi_\lambda \ket =0$. Hence, in taking the derivative with respect to the parameter $\lambda$ in the path-integral expression, we can ignore its action on the asymptotic boundary conditions and write
$$
\pt_\lambda E_n (\lambda) = {1\over {\rm Vol} ({\bf R})} \int [\CD \phi]_n \, \pt_\lambda S_\lambda \,e^{-S_\lambda} \;.
$$
We now generalize the leading order deformation in $\lambda$ and define the local operator $\CO_\lambda$ as the one satisfying 
$$
\pt_\lambda S_\lambda = \int dt\, \CO_\lambda (t)\;.
$$
In other words, $\CO_\lambda$ defines the infinitesimal deformation $\lambda \rightarrow \lambda + \delta \lambda$ for any finite value of $\lambda$. In particular, the operator $\CO_\lambda$ generally depends on $\lambda$. 
We can regard the deformation as `local' if  $\CO_\lambda (t)$ is a  polynomial in  $\phi(t)$ and derivatives of $\phi(t)$. 

With these definitions we obtain the following spectral flow equation
\eqn\flowone{
\pt_\lambda E_n (\lambda) = \bra n | \CO_\lambda | n\ket\;.
}
The disappearance of the volume factor follows from
$$
\int dt \bra n | \CO_\lambda (t) | n\ket = \int dt e^{iE_n (\lambda)  t} \bra n| \CO_\lambda (0) |n\ket e^{-itE_n (\lambda)} = {\rm Vol} ({\bf R}) \bra n | \CO_\lambda (0) | n \ket \;.
$$

A simple example of the form \flowone\ is obtained by adding $ {\cal O}_0 =- \ell\,H^2$ to the  action of a given quantum-mechanical model. The  iterated `quadratic' flow is given by the solution of 
\eqn\qua{
\partial_\ell E_n (\ell) = -E_n (\ell)^2\;,
}
which integrates to 
\eqn\hdosdef{
F(E) = {E \over 1+  \ell E}\;.
}
In describing this  example we shall use the coupling $\ell$, with dimensions of length, as the deformation parameter. 
For small deformations, the quadratic  flow \qua\  is pathological at high energies when $\ell<0$, due to the pole at $E_{\rm pole} =1/| \ell |$. For $\ell >0$, there is no urgent high-energy pathology, but there  is  potential low-energy trouble if the ground state energy $E_0$ is negative, in which case we must limit the deformation to be sufficiently small $\ell < 1/|E_0|$. These properties are qualitatively similar to those of the $T{\bar T}$ flow of $d=2$. 
The main difference is the rather radical UV squeezing for $\ell >0$: the whole spectrum has a UV accumulation point, or absolute UV cutoff at $\Lambda_F = 1/\ell$, suggesting a very large density of states near this accumulation point. 

Results such as  \hdosdef\  touch upon the following point.
One starts off by adding to the Lagrangian  an irrelevant operator, such as $-\ell\, H^2$. The resulting Lagrangian exhibits higher time derivatives, so that  it does not lend itself to a straightforward canonical quantization and in particular not to the exact form of the modified Hamiltonian. On the other hand,  using the finite deformation versions \flowone\  and \qua, one does obtain the exact form of the deformed Hamiltonian such as \hdosdef.   An expansion in terms of  a small deformation parameters allows to identify an approximate Hamiltonian 
for that case.

At any rate, the flexibility of quantum mechanics allows us to consider very general deformations. For example, we may generalize the quadratic flow \qua\ to a `fractional' deformation of the form
\eqn\deltadef{
\partial_\lambda E_n (\lambda) = -{1\over \delta} \,E_n(\lambda)^{1+\delta}
}
with non-integer $\delta$. For $\delta>0$ we get fractional deformations with similar properties as \hdosdef\ above, whereas $\delta <0$ gives a much milder type of deformation with little impact on the UV spectral properties. 

In looking for $d=1$ analogs of the  $T{\bar T}$ deformation, we may emphasize different aspects of the $d=2$ deformation. The deformation \hdosdef\ emphasizes the quadratic form of the flow equation \qua. Alternatively, we may follow \refs\gross\ and  just write down the same final formula that we obtained in $d=2$, 
\eqn\ttbar{
F(E) = {1\over 2 \ell} \left(\sqrt{1+4\ell E} -1\right)\;,
}
which, by construction, has the same formal properties as the $d=2$ case. In this case the flow equation 
$$
\pt_\ell E_n (\ell) = -{ E_n (\ell)^2 \over 1+ 2 \ell E_n (\ell)}
$$
differs from \qua\ by higher order terms in the control parameter $\ell$. More generally, given any regular invertible function $f(x)$ with non-vanishing derivative $f'(x)$, the formal flow equation
\eqn\genflow{
\pt_\lambda H_\lambda =  f' (f^{-1} (H_\lambda))
}
may be solved by
\eqn\genflowe{
F= H_\lambda = f \left(\lambda + f^{-1}(E)\right)
\;,
}
where $f^{-1}$ is the inverse function. A crucial property of all these QM deformations $H \rightarrow F(H)$ is the fact that the energy eigenstates remain formally untouched. 

Regarding the seed theory, we also have considerable freedom. Natural examples to consider are bosonic particles  in various potentials, such as a harmonic trap, or a box in various dimensions. On the other hand, nothing prevents us from doing a Hamiltonian deformation of a many-body theory in arbitrary dimensions, such as a $d$-dimensional field theory but, in such cases, the deformed theory is radically non-local for $d>1$.

Focusing on the $d=1$ realm, any  particle model with a discrete spectrum has a characteristic high-energy microcanonical temperature,
\eqn\miuni{
T_0 (E) =  b\, E + \dots\;,
}
where the dots stand for subleading terms in the large-$E$ limit and $b$ is a model-dependent constant (see Appendix C for a derivation of this fact).  Considering for instance  a particle in a $D$-dimensional potential, $b= 1/D$ for a harmonic trap and $b=2/D$ for a sharp, spherical box. From this point of view, the target-space dimensionality of the QM model plays a similar role as the central charge in  $d=2$.

The microcanonical temperature of the deformed theory is given by equation \mict, with the high energy asymptotics
$$
T(F(E)) \approx F'(E) \, b \, E\;,
$$
so that any $F'(E)$ decaying faster than $1/E$ ensures a positive specific heat. In other words, any deformation  function growing faster than logarithmically will leave the deformed theory thermodynamically stable. This is the case for the deformation \ttbar, but not for the quadratic one \hdosdef. The quadratic deformation being asymptotic to a constant, induces a monotonically decreasing temperature function for the deformed theory. 
More precisely, we have 
\eqn\fint{
T(F) \approx b F (1-\ell F) \;,
}
an inverted parabola with a maximum at $F_s =1/2\ell$ and negative specific heat in the interval $1/2\ell < F < 1/\ell$, a form qualitatively similar to that of asymptotically flat black holes. 

This qualitative behavior generalizes to all the $\delta$-deformations \deltadef\  with $\delta >0$,
\eqn\deltadeff{
F(E)= {E \over  (1+\lambda E^\delta)^{1/\delta}} \;, 
}
 for which the deformed theory has the full spectrum accumulated below the maximum $\Lambda_F = (1/\lambda)^{1/\delta} $ and \fint\ generalizes to
 \eqn\findelta{
 T(F) \approx bF (1-\lambda F^\delta)\;.
 } 
 This curve has a maximum at $F_s = [(1+\delta)\lambda]^{-1/\delta}$ and vanishes linearly with derivative $-b\, \delta$ as we approach the endpoint energy. Hence, beyond $F_c$ the system has negative specific heat. 
 
 It is interesting to determine the critical deformation with the borderline behavior, namely the Hagedorn case. A constant $T(F)$ requires $F'(E) \sim 1/E$ at high energy, which fixes the logarithmic growth for the function $F(E)$. If we further require that  the flow equation be \qua\ at leading order in the deformation parameter, we have
 \eqn\hagc{
 F(E) = {1\over 2\ell} \log (1+2\ell E) \;,
 } 
 which leads to a Hagedorn temperature $T_s = b/2\ell$ in the deformed theory. Again, we notice that $T_s \ll 1/\ell$ when the `central charge' is large, since $b \sim 1/D$. 
  The corrections which determine the sign of the specific heat depend on details, as can be seen by examining the two simplest examples. 
 
 Consider  a harmonic oscillator with spectrum $E_n = \omega \,n$ (after  shifing the vacuum energy to zero, with $n$ a non-negative integer),  the level counting function and  the associated microcanonical temperature read 
 $$
 {\cal N}_0 (E) = 1+ {E \over \omega} \;,\qquad
 T_0 (E) = E + \omega\;.
 $$
Using then the basic equation \mict\ one obtains the microcanonical temperature of the Hagedorn-defomed theory:
\eqn\ftt{
T(F) = {1\over 2\ell} + \left(\omega -{1\over 2\ell}\right) \,e^{-2\ell F} \;.
}
This function increases monotonically towards the Hagedorn temperature $1/2\ell$ for small $\ell$, This behavior turns over for $\ell > 1/2\omega$, in which case the specific heat becomes negative. 

Another simple example worth considering is a particle of mass $m$ going around a circle of radius $R$. In this case, the  non-vacuum spectrum is two-fold degenerate, $E_n = n^2 / 2m R^2$, with a level function and microcanonical temperature 
$$
{\cal N}_0 (E) = 1+ R\sqrt{8mE}\;, \qquad 
T_0 (E) = 2E \left(1+ {1 \over R\sqrt{8mE}}\right)\;,
$$
leading to a Hagedorn-deformed temperature function: 
\eqn\defff{
T(F) = {2E(F) \over 1+ 2\ell E(F)} \,\left(1+ {1 \over R\sqrt{8mE(F)}}\right) = {1\over \ell} \left(1+ \sqrt{\ell  \over 4mR^2} \;e^{-\ell F}  + O\left(e^{-2\ell F}\right)\right)\;,
}
which is monotonically decreasing at high energy, signaling a negative specific heat. Hence, we see that small differences in otherwise natural examples are capable of tipping  the  thermodynamic stability of the Hagedorn spectrum one way or another. 

We end this section by highlighting some special features which emerge for systems that
are in the framework of those which were analyzed above.
In Section 1 we discussed how symmetries that manifest themselves by leading to 
degeneracy among energy levels in the seed system metamorphose but are retained 
in the deformed system. 
There are however symmetries which manifest in different ways,
an example of such a symmetry is the $SL(2,{\bf R})$ symmetry of the  conformal \refs\fubini\ 
and superconformal QM \refs{\fubinir, \ivanov}. In those cases, a  deformation such as \hdosdef\  explicitly breaks the scale
symmetry and this does manifest in the deformed symmetry by  having an energy spectrum cutoff
at energy $1/\ell$.  However the introduction of that explicit scale is not enough to provide a normalizable zero energy ground state for these systems; the wave functions remain unchanged. That signature of the presence of scale invariance persists as do various symmetries present in the supersymmetric case. 

Another aspect of interest is the study of the fate of bound states under the deformation.
In particular consider the deformation in Eq. \hdosdef\ 
for the case of the hydrogen atom.
For bound states $E_n = -E_R / n^2$  one obtains
\eqn\hdosdeff{
F(E_n) = {-E_R  \over n^2 -  \ell E_R}\;.
}
When the deformation parameter $\ell$ is negative, as far as the bound states are concerned, 
their energies accumulate at zero as in the seed theory. The positive energies will hit a pole at
at the value $1/|\ell |$. 
However things become more exotic for positive values of $\ell$.
If $\ell E_R $ happens to have a value equal to a square of an integer $j$, the deformed energy of
the $j$th bound state will become unbounded from below, a negative infinity,  and the system
will be unstable. If $\ell E_R$ is not of that form then bound states whose value of $n$ is less than
$\sqrt{\ell E_R}$ will have a positive energy and, as they all do retain their normalizability property, as 
well as their $n^2$ degeneracy, they will be embedded in the continuum. The larger $\ell$ is, the 
more positive bound states embedded in the continuum there will be. That could be a very
interesting condensed matter system.

\newsec{Discussion}

\noindent

In this brief note we have digressed over whether $T{\bar T}$ deformations represent a challenge to Wilsonian wisdom, namely the expectation that a single irrelevant operator will not be enough to completely reconstruct the UV theory from the IR theory. In this respect, $T{\bar T}$ deformations represent an interesting middle ground, being non-trivial but still exactly solvable. We have focused on the high-energy behavior of various examples, from the original $T{\bar T}$-deformed CFT in two dimensions, to quantum mechanical analogs. When the high-energy spectrum is free from blatant instabilities, such as formally complex energies, we notice that there is a tendency for the deformed spectra to accumulate at high-energies, increasing the density of states. The benchmark model in $d=2$ is a marginal case, with a Hagedorn density of states, which we show to have negative specific heat, signaling a generic instability if the theory is embedded into any UV completion with more standard thermodynamics. For the quantum mechanical models, the effect is more drastic, as the UV accumulation of the spectrum is quite literal with the emergence of an absolute energy cutoff and a density of states qualitatively similar to that of black holes in flat space. 

An interesting question is whether the models with strong thermodynamical instability in the UV are `physically' acceptable at all. In particular, what would be the features of a system which has at high energy a negative specific heat, what would be the price for that.
A possible  
 answer is that they are legal, provided we give up any pragmatic notion of locality \refs\banksofer. 
In standard Wilsonian theories we have a CFT governing the UV and local operators have correlation functions defined
for any non-zero time separation.  In particular, for operators of definite scaling dimension $\Delta$ we have
\eqn\coo{
\bra \CO(t) \CO(0) \ket = {1\over |t|^{2\Delta}}\;.
}
Introducing a spectral representation we can write  
\eqn\spec{
\bra \CO(t) \CO(0) \ket = \int_0^\infty  dE \,\Omega_0 (E) |\bra E | \CO | 0 \ket |^2 \,e^{-iEt} 
}
in terms of the density of states $\Omega_0 (E) = d {\cal N}_0 / dE $. The sub-exponential growth of $\Omega_0 (E)$, characteristic of  any QFT,   ensures that the correlation is analytic as $t\rightarrow t- i \epsilon \cdot {\rm sign}(t)$, so that we can define it by analytic continuation from the Euclidean counterpart. This breaks down for densities of states with  Hagedorn asymptotics or harder, such as black holes. Therefore, we expect that time locality will accordingly break down for those quantum mechanical models
with Hageorn-like   deformations. 
The case of deformed QM  with bounded maximal energy  is interesting, since the integral over energies is now cutoff by $
\Lambda_F$,  and one may wonder if local time correlations can be defined after all. 

To check this, we begin by noticing that any deformation $H \rightarrow F(H)$ will conserve the energy eigenstates. Therefore, if we keep the operator fixed under the flow, it will have the same matrix elements in the energy basis, 
$
\CO_{mn} = \bra n | \CO | m \ket \,,
$
before and after the flow. For scaling operators of definite dimension $\Delta$, dimensional  analysis  implies that the $t^{-2\Delta}$
behavior of \spec\   is equivalent to 
$
\bra E | \CO | 0 \ket \sim E^\Delta
$. Here we use the fact that $\Omega_0 (E) = d{\cal N}_0 (E) /dE \sim 1/E$ at high energies in QM systems. 
Therefore,  we can write  
$
\bra F | \CO | 0\ket = E(F)^{\Delta} 
$
when expressed in terms of the new energies. If the deformation has a UV accumulation point, as in the case of the standard quadratic flow, 
\eqn\otravez{
F(E) = {E \over 1+ \ell E} \;,
}
with $F_{\rm max} = \Lambda_F = 1/\ell$, 
the correlation function at finite $\ell$ becomes
\eqn\almost{
\bra \CO(t) \CO(0)\ket_\ell = \int_0^{\Lambda_F} dF \,\Omega(F) \,|\bra F | \CO | 0\ket |^2 \,e^{-iFt} = \int_0^\infty dE \,\Omega_0 (E) \,|\bra E | \CO | 0\ket |^2 \,e^{-i F(E) t}
\;.
}
As $E\rightarrow \infty$ the time-dependent phase stops oscillating due to the accumulation point in $F$, so that the correlation function
has a UV  contribution at any {\it finite} $t$ given by 
\eqn\there{
e^{-it/\ell}  \int^\infty dE \,\Omega_0 (E) \,E^{2\Delta} \sim e^{-it/\ell} \int^\infty {dE \over E} \,E^{2\Delta} \;,
}
which diverges for any $\Delta \geq 0$. Hence, only operators with $\Delta <0$ seem to define correlation functions at arbitrarily small times, but then this would violate unitarity bounds, and the answer is regular in $t$, with no scaling properties.  These considerations generalize verbatim to the positive delta-deformations \deltadef\ which share the qualitative properties of the quadratic deformation. 

Hence, we conclude that the simplest analogs of $T{\bar T}$-deformation in quantum mechanics, featuring spectral accumulation points, tend to change the UV in a way that jeopardizes locality, even in its  weakest forms. 

\vskip1cm

\centerline{\bf Acknowledgements} 

We wish to thank Sasha Zamolodchikov for his lectures at the Jerusalem winter school of 2018 which inspired  this note. We also would like to thank Ofer Aharony and Thomas Banks for comments on the manuscript. E. Rabinovici would like to thank NHETC at Rutgers Physics Department and CCPP at NYU for hospitality, and Sergey Dubovsky, David Kutasov, Matt Kleban and Ruth Shir for discussions. 
J.L.F. Barbon would like to thank the Hebrew University in Jerusalem, where most of this work was done and the IHES in Bures sur Yvette for hospitality. The work of E. Rabinovici is partially supported by the Israeli Science Foundation Center of
Excellence.  
 The work of J.L.F. Barbon  is partially supported by the Spanish Research Agency (Agencia Estatal de Investigaci\'on) through the grants IFT Centro de Excelencia Severo Ochoa SEV-2016-0597,  FPA2015-65480-P and 
PGC2018-095976-B-C21. 



\appendix{A}{ Logarithmic Corrections To Cardy's Formula}

\noindent

The high-energy density of states has a rather universal behavior for two-dimensional  CFTs. In particular, this universality extends to the  corrections to Cardy's classic result.  To discuss this, we can evaluate the density of states by inverting the expression for the modular invariant partition function:
\eqn\direct{
Z(\tau, {\bar \tau}) = \Tr q^{\;L_0 - c/24} \;{\bar q}^{\;{\bar L}_0 - {\bar c}/24} = \int d\varepsilon \,d{\bar \varepsilon}  \;\rho (\varepsilon, {\bar \varepsilon}) \;e^{2\pi i \tau \varepsilon - 2\pi i {\bar \tau} {\bar \varepsilon} }   
\;,
}
where $q= e^{2\pi i \tau}$,  ${\bar q} = e^{-2\pi i {\bar \tau}}$ and 
\eqn\denc{
\rho (\varepsilon, {\bar \varepsilon}) = \sum_\alpha \delta (\varepsilon - \varepsilon_\alpha) \,\delta({\bar \varepsilon}- {\bar \varepsilon}_\alpha) 
}
denotes the density of states as a function of the `chiral energies', related to the conformal dimensions $(\Delta, {\bar \Delta})$  by the expressions 
$\varepsilon \equiv \Delta - c/24$ and ${\bar \varepsilon} \equiv {\bar \Delta} - {\bar c}/24$. The standard energy and momentum of the theory on a circle of unit radius are    $E= \varepsilon + {\bar \varepsilon} $ and $P = \Delta-{\bar \Delta}= \varepsilon - {\bar \varepsilon} + (c-{\bar c})/24$. This  change of variables determines the relation to the density of states as a function of energy and momentum:
$$
\rho(\varepsilon, {\bar \varepsilon}) \,d\varepsilon \,d{\bar \varepsilon} = \Omega(E, P) \,dE\,dP\;,
$$
 which gives $\rho (\varepsilon, {\bar \varepsilon}) = 2\, \Omega (E, P)$. Therefore, the inversion formula for the density of states as a function of energy and momentum is given by
\eqn\denep{
\Omega (E, P) = {1\over 2}\, \rho (\varepsilon, {\bar \varepsilon}) = {1\over 2}\, \int_{\Gamma} d\tau \,e^{-2\pi i \tau \varepsilon} \int_{\bar \Gamma} d{\bar \tau} \,e^{ 2\pi i {\bar \tau} {\bar \varepsilon}} Z(\tau, {\bar \tau})\;.
}
In this formula $\tau$ and ${\bar \tau}$ are regarded as independent integration variables. The contour $\Gamma$ runs parallel to the real $\tau$ axis, with positive ${\rm Im} \,\tau$ large enough to lie entirely on the analyticity domain of the partition function. The contour ${\bar \Gamma}$ lies similarly at a fixed and sufficiently negative value of  ${\rm Im} \,{\bar \tau}$. 

Following \refs\carlip, it is convenient to rewrite the partition function as 
\eqn\modi{
Z(\tau, {\bar \tau}) = q^{-{c_{\rm eff} \over 24}} \,{\bar q}^{-{{\bar c}_{\rm eff} \over 24}} \,{\widetilde Z} (\tau, {\bar \tau})\;,
}
where $c_{\rm eff} = c- 24 \Delta_0$, ${\bar c}_{\rm eff} = {\bar c} - 24{\bar \Delta}_0 $ and  $(\Delta_0, {\bar \Delta}_0)$ are the lowest conformal dimensions.  The renormalized partition function
\eqn\moddi{
{\widetilde Z} (\tau, {\bar \tau}) = \Tr \,q^{L_0 - \Delta_0} \; {\bar q}^{\;{\bar L}_0 - {\bar \Delta}_0}  = \sum_\alpha e^{2\pi i \tau (\Delta_\alpha - \Delta_0) - 2\pi i {\bar \tau} ({\bar \Delta}_\alpha - {\bar \Delta}_0)} 
}
has the crucial property that, for unitary CFTs with real and discrete spectrum of conformal dimensions, it approaches a constant with  exponential accuracy  as ${\rm Im}\,\tau \rightarrow \infty$ and ${\rm Im}\,{\bar \tau} \rightarrow -\infty$. This limiting constant is ${\cal N}(0)$, i.e. the degeneracy of the ground state. 

Under these circumstances, we can use the modular invariance of the partition function, $Z(\tau, {\bar \tau}) = Z(-1/\tau, -1/{\bar \tau})$, to rewrite \denep\ in the form
\eqn\newdenep{
\rho (\epsilon, {\bar \varepsilon}) =  \int d\tau\; d{\bar \tau} \; e^{-2\pi i \tau \varepsilon +{2\pi i c_{\rm eff} \over 24 \tau}} \;e^{2\pi i {\bar \tau} {\bar \varepsilon} - {2\pi i {\bar c}_{\rm eff} \over 24 {\bar \tau}}} \; {\widetilde Z} (-1/\tau, -1/{\bar \tau})\;.
}
Now, in the limit that $\varepsilon$ and ${\bar \varepsilon}$ are large, this integral is dominated by saddle points 
\eqn\sad{
\tau_s = i\sqrt{c_{\rm eff} \over 24\, \varepsilon} \;, \qquad {\bar \tau}_s = -i \sqrt{{\bar c}_{\rm eff} \over 24\, {\bar \varepsilon}} \;,
}
which sit in the region where ${\widetilde Z} (-1/\tau, -1/{\bar \tau})$ is exponentially close to the constant ${\cal N}(0)$. Evaluating both integrals in the gaussian approximation one finds 
\eqn\fff{
\rho (\varepsilon, {\bar \varepsilon}) \approx {\cal N}(0) \,c_{\rm  eff}\, {\bar c}_{\rm eff} \;{\pi^3 \over 18}  \,{e^{S_L + S_R} \over \left(S_L \, S_R \right)^{3/2}}   \;,
}
up to corrections in inverse powers of $S_L, S_R$, 
where 
\eqn\ls{
S_L (\varepsilon) = 2\pi \sqrt{{c_{\rm eff} \over 6} \,\varepsilon} \;, \qquad S_R ({\bar \varepsilon}) = 2\pi \sqrt{{{\bar c}_{\rm eff} \over 6} \,{\bar \varepsilon}}  \;.
}
 It is interesting to remark that, for CFTs with integer-spaced conformal dimensions, it is possible to sum all power corrections into 
\eqn\fffe{
\rho (\varepsilon, {\bar \varepsilon}) \approx {\cal N}(0) \,{\pi^4 \over 9} \, c_{\rm eff} \,{I_1 (S_L) \over S_L} \; {\bar c}_{\rm eff} \,{I_1 (S_R) \over S_R}  \;,
}
where $I_1 (x)$ is a modified Bessel function, the remaining corrections being exponentially suppressed in $S_{L,R}$ (cf.  \refs\sheikh). 

From \fff\ we can deduce the density of states in the undeformed theory, projected to the zero-momentum sector and finally the critical exponent $\gamma$ in \cardy. Focusing on left-right symmetric CFTs with $c = {\bar c}$ we find 
\eqn\porj{
\Omega_0 (E) \Big |_{P=0} = \Omega_0 (E, 0) =  {2\pi^3 c^2 \over 9}  {e^{S_C (E) } \over S_C(E)^3} \left(1+ O(1/S_C)\right)\;,
}
with $S_C(E) = 2\pi \sqrt{ cE/3}$ the Cardy entropy. In \porj\ and below, we simplify the notation by  writing $c$ in place of $c_{\rm eff}$, with the understanding that we actually mean $c_{\rm eff}$ whenever the two are different.

In order to compute the microcanonical temperature we need to integrate \porj\ once with respect to the energy to obtain
\eqn\ncerom{
{\cal N}_0 (E) \big |_{P=0} = {\pi c \over 3} \,{e^{S_C (E)} \over S_C (E)^2} \left(1+ O(1/S_C)\right) \;,
}
which leads to the microcanonical entropy  with $\gamma \big |_{P=0} =2$, 
\eqn\microen{
S_0 (E) \big |_{P=0} = S_C (E) - 2 \,\log S_C (E) + \log (\pi c / 3) + O(1/S_C) \;,
}
and finally  the microcanonical temperature quoted in \cardlog. Incidentally, the same methods can be used to compute the full density of states as a function of the energy, after integrating over momenta, $\Omega_0 (E) = \int dP\, \Omega_0 (E, P)$. The most efficient way of doing this is to go back to
\direct\ and set $\tau = i\tau_2$ to be pure imaginary. Then one replaces \denep\ by the standard Laplace inversion formula, with a single contour integral, and finally obtains $\delta = 3/2$ and  $\gamma = 1/2$, in the notation of equations \genp\ and \genn.

It is interesting to discuss how these results are modified when the requirement of `compactness' of the CFT is lifted.\foot{We thank O. Aharony for discussions on this issue.} After all, the simplest example of a $T{\bar T}$ deformation starts from a free boson, which has a continuous spectrum of conformal dimensions. We can discuss this particular example by regularizing the theory as a free compact boson with radius $r$. More generally, we can consider a CFT which is the direct product of a compact CFT of central charge $c'$ and $N$ such bosons, which contribute a sigma model factor whose target space is a straight torus of volume $(2\pi r)^N$. The total central charge is $c=c' +N$ and the modular-invariant partition function is given by 
\eqn\freb{
Z(\tau, {\bar \tau}) = \left[{r \over \sqrt{2}}  {({\rm Im} \,\tau)^{-1/2}  \over | \eta (\tau) |^2 }\right]^N \cdot Z' (\tau, {\bar \tau})\;
}
in the $r\rightarrow \infty$ limit (cf. for instance \refs\difran).   Here $\eta (\tau) = q^{1/24} \prod_{n=0}^\infty (1-q^n) $ is Dedekind's eta function and $Z'(\tau, {\bar\tau})$ is the partition function of the compact CFT. Since the total partition function is proportional to the divergent volume of the target space, $|{\bf T}_N | = (2\pi r)^N$,  the resulting density of states will be defined per unit volume of this target space. 

Following the same strategy as in the compact case, we notice that the function
\eqn\newtil{
{\widetilde Z}(\tau, {\bar \tau}) = {{\widetilde Z}' (\tau, {\bar \tau}) \over | \varphi (\tau)|^{2N}}\;,
 \qquad {\rm with} \;\;\varphi (\tau) \equiv q^{-1/24} \,\eta(\tau)\;,
 }  
 has the  property of approaching a constant  exponentially fast, as ${\rm Im} \,\tau \rightarrow \infty$. In this case this constant is ${\cal N}'(0)$, the vacuum degeneracy of the compact CFT factor. 
 
 We can now write an inversion formula analogous to \newdenep:  
\eqn\ncdenep{
\rho(\varepsilon, {\bar \varepsilon}) = {|{\bf T}_N |  \over \left(\pi \sqrt{8}\right)^N}  \int d\tau\; d{\bar \tau} \; e^{-2\pi i \tau \varepsilon +{2\pi i c \over 24 \tau}} \;e^{2\pi i {\bar \tau} {\bar \varepsilon} - {2\pi i {\bar c} \over 24 {\bar \tau}}} \;\left({2i \tau {\bar \tau} \over (\tau - {\bar \tau})}\right)^{N\over 2} {\widetilde Z} (-1/\tau, -1/{\bar \tau})\;.
}
The rational power  in the integrand comes from the modular transformation of $({\rm Im}\,\tau)^{-N/2}$, written  in terms of $\tau$ and ${\bar \tau}$, in order to make explicit the fact that the integration in \ncdenep\ treats these two variables as independent. 
In the limit of large chiral energies, this integral is dominated by the same saddle points \sad, up to small corrections induced by the power term in the integrand. Explicit evaluation in the gaussian approximation yields a modification of \fffe\ which amounts to the substitution 
\eqn\su{
{\cal N}(0) \longrightarrow {\cal N}' (0) \, {|{\bf T}_N | \over \left(\pi \sqrt{8}\right)^N}  \;\left({2i \tau_s {\bar \tau}_s \over (\tau_s - {\bar \tau}_s)}\right)^{N\over 2} \;,
}
with $\tau_s$ and ${\bar \tau}_s$ given by the expressions \sad. Upon explicit calculation, we obtain the final substitution rule 
\eqn\subb{
{\cal N} (0) \longrightarrow  {\cal N}' (0)\; {|{\bf T}_N | } \left({c \over 48 \pi}\right)^{N/2} {1 \over (S_L + S_R)^{N/2}} \;.
}
This means that the critical exponent controlling the $ P=0$ specific heat of the deformed theory is now given by
\eqn\newc{
\gamma \big |_{P=0} = 2 + {N \over 2} \;,
}
a result which does not change the thermodynamical stability properties of the deformed theory.

\appendix{B}{Estimating The Sign Of The Full  Heat Capacity}

\noindent

For simplicity, our analysis of the thermodynamic  stability  in section 1 was restricted to the zero-momentum sector. As remarked in \fulls, the spectral `squeezing' is less pronounced in the $P\neq 0$ sectors. Therefore, we expect that the negativity of the heat capacity will be maintained when the finite-momentum sectors are summed up in the computation of the density of states. In this appendix we provide a formal argument supporting this expectation. 

Given the density of states in the deformed theory as a function of energy and momentum, $\Omega (F, P)$, we can obtain the integrated density which depends only on energies as 
\eqn\intee{
\Omega (F) = \int_{P_-}^{P_+} dP \,\Omega (F, P)\;,
}
where the endpoints $P_\pm$  in general depend on $F$ and  satisfy $P_- <0<P_+$ and $|P_+| = |P_-|$. At high undeformed energies $E\gg 1$, in units of the circle's radius ($R=1$),  we have $P_\pm \approx \pm E$.  Substituting this into  the inverted form of \fulls\  at general momentum,
\eqn\inverted{
E = F + 2\pi \lambda F^2 - 2\pi \lambda P^2 = F + {F^2 \over F_s} - {P^2 \over F_s}\;,
}
where we recall the definition of the critical energy $F_s = 1/ 2\pi \lambda$, in units $R=1$. Solving for $P_\pm$ in the Hagedorn regime $F\gg F_s$, we find  $P_\pm \approx \pm F$. 

The function $\Omega (F)$ can be related  to the undeformed densities of states as follows. The general expression for the deformed density is
$$
\Omega (F, P) = \sum_\alpha \delta(F-F_\alpha) \,\delta (P-P_\alpha) \;,
$$
where $F_\alpha$ and $P_\alpha$ are the actual discrete values of the energy and momenta in the state $|\alpha\ket$. Changing variables to undeformed energies we can write
$$
\Omega (F(E, P), P) = \sum_\alpha {1\over | \pt_E F |} \delta (E-E_\alpha) \,\delta (P-P_\alpha)\;.
$$
By explicit computation, 
$$
\pt_E F = {1\over 1+4\pi \lambda F}\;,
$$
which leads to 
\eqn\relaun{
\Omega (F, P) = (1+ 4\pi \lambda F) \,\sum_\alpha \delta(E(F,P)-E_\alpha) \,\delta(P-P_\alpha) = (1+4\pi\lambda F) \Omega_0 (E(F,P), P)\;,
}
in terms of the undeformed density of states 
$$
\Omega_0 (E,P) = \sum_\alpha \delta (E-E_\alpha) \,\delta (P- P_\alpha) \;,
$$
 whose leading high-energy behavior scales as  
 \eqn\undeff{
 \Omega_0 (E, P) \propto  {1\over \left(S_C (E+P) + S_C (E-P)\right)^{N/2} } \;{e^{S_C (E+P)/2} \over S_C (E+P)^{3/2}} \,{e^{S_C (E-P)/2} \over S_C (E-P)^{3/2}} + \dots 
 }
 In this expression, we have  written $S_L$ and $S_R$ from \fff\ in terms of $S_C$ and kept only the leading and next-to-leading terms at large energies. Furthermore, in addition to the general scaling from \fff, we have
 included a term of the form \subb\ to incorporate the possible occurrence of non-compact factors, such as  $N$ free bosons. 
 
 The resulting density of states of the deformed theory reads 
 \eqn\denden{
 \Omega (F, P)  \propto (1 + 2 F /F_s) \,e^{h(F, P)} \;, \qquad h(F,P) \equiv \log \,\Omega_0 \left( E(F, P), P\right)\;.
 }
 The function $h(F, P) =f(E(F,P),P)$ is even in $P$ and maximal at $P=0$. This justifies the approximation of the density of states by the maximum at zero momentum. In order to estimate the quantitative effect of actually integrating over all momenta, we evaluate \intee\ in the saddle-point approximation:
\eqn\sadd{
\Omega(F) \propto  \left(1+ 2 {F \over F_s}\right) \,\sqrt{-{2\pi \over h_{PP} (F) }} \, {e^{S_C (E(F, 0))}  \over S_C (E(F, 0))^{3+{N\over 2}}} \left(1+ O(1/S_C) \right)\;,
}
where $h_{PP}(F)$ is the second derivative
$$
h_{PP} (F) = {\pt^2 \over \pt P^2} \;h(F, P) \Big |_{P=0} \;.
$$
Recalling that $S_C (E(F)) \approx F/T_s$ as $F\gg F_s$, direct evaluation of this derivative in the Hagedorn regime yields 
$$
h_{PP}  (F) \approx  -{4\pi^2 c \over 3} {T_s \over F_s } {1\over F}\;.
$$
This means that the width of the distribution around the maximum at $P=0$ is proportional to  $\sqrt{F}$, smaller than the support of the integral \intee, which is of order $2|P_+ | \sim F$. Under these circumstances, we can neglect the corrections to \sadd\ coming from the cut tails and conclude that 
\eqn\omef{
\Omega (F) \propto  F^{3/2}\; {e^{F/T_s} \over F^{3+{N \over 2}}} + \dots\;.
}
Integrating once with respect to $F$ we obtain the deformed level-counting function 
\eqn\defl{
{\cal N} (F) \propto {e^{F/T_s} \over F^{3+N \over 2} } + \dots 
}
which leads to an asymptotic  microcanonical temperature 
\eqn\otraf{
T(F) = T_s \left( 1+{3+N\over 2}\,{T_s \over F} + \dots \right) \;,
}
with negative heat capacity. 


\appendix{C}{WKB approximation to the microcanonical temperature in QM}

\noindent

The high-energy asymptotic behavior of the microcanonical temperature in QM, $T(E) \approx b E$, can be established in the WKB approximation, which should be justified precisely at high excitation levels, when sums over states can be approximated by integrals. Let us consider  a particle in $D$ dimensions with Hamiltonian
\eqn\hamim{
H({\vec p}, {\vec q}\;) = {{\vec p}^{\;2} \over 2m} + V({\vec q}\;)\;,
}
whose potential ensures bounded motion at fixed energy and thus a discrete spectrum. 
 The WKB formula for the density of states reads (we set $\hbar =1$) 
 \eqn\wkbomega{
\Omega (E) \approx \int {d^D p \;d^D q \over (2\pi)^D} \,\delta\left( E - H({\vec p}, {\vec q}\;)\right) = m {|{\bf S}^{D-1} | \over (2\pi)^D} \int d^D q \;\left[2m(E-V({\vec q}\;)\right]^{D-2 \over 2} \;,
}
where  $|{\bf S}^{D-1}| $ is the volume of the corresponding unit sphere and we have evaluated the momentum integral in the last equality. Further assuming spherical symmetry we have
\eqn\wkbsphe{
\Omega (E) \approx m {|{\bf S}^{D-1} |^2 \over (2\pi)^D} \int q^{D-1} \,dq\,\left[2m(E-V(q))\right]^{D-2 \over 2}\;.
}
Extracting a power of $E^{D-2 \over 2}$ from the integral, the dependence on the potential is proportional to $V(q)/E$, thus  the value of the potential away from the `turning points' $V(q_E) =E$ is not relevant in the large-$E$ limit. This suggests that the qualitative behavior is the same as that of a free particle in a box, which indeed has $T(E) \propto E$. 

The proportionality coefficient depends on the details of the potential. If $V(q)$ is polynomial with a highest power $V(q) \sim q^n$, a rescaling of the integration variable by $q \rightarrow E^{1/n} \,q$ shows that 
$$\Omega (E) \propto E^{\,a} \;, \qquad a= {D-2 \over 2} + {D \over n}\;.
$$
Hence, the entropy scales like $S(E) \sim (a+1)\,\log (E)$ and the microcanonical temperature reads
\eqn\miccc{
T(E) \approx b\, E\;, \qquad b = {1\over a+1} = {2n \over (n+2)D}\;.
}
We find $b=1/D$ for the harmonic oscillator and $b= 2/D$ for the sharp spherical box, which is obtained in the formal $n\rightarrow \infty$ limit.

{\ninerm{
\listrefs
}}

\end